\documentclass[twocolumn,pra,floatfix]{revtex4-1} 

\usepackage{graphicx}
\usepackage{amsmath, amssymb, color}

\allowdisplaybreaks

\newcommand{\ktr}{k_{\rm tr}}

\begin{document}
 
\title{Bose-Hubbard ladder subject to effective magnetic field: 
	quench dynamics in a harmonic trap}

\author{Wladimir Tschischik}
\author{Roderich Moessner}
\author{Masudul Haque}

\affiliation{Max-Planck-Institut f\"ur Physik komplexer Systeme, N\"othnitzer
Strasse 38, 01187  Dresden, Germany} 

\date{\today}

\begin{abstract}

Motivated by a recent experiment with optical lattices that has realized a 
ladder geometry with an effective magnetic field (Atala et al., Nature Physics 
\textbf{10}, 588 (2014)), we study the dynamics of bosons on a tight-binding 
two-leg ladder with complex hopping amplitudes.  This system displays a quantum 
phase transition even without interactions.  We study the non-equilibrium
dynamics without and with interactions, in the presence of a harmonic trapping 
potential.  In particular we consider dynamics induced by quenches of the 
trapping potential and of the magnitude of the rung hopping.  We present a 
striking "slowing down" effect in the collective mode dynamics near the phase 
transition.  This manifestation of a slowing down phenomenon near a quantum 
phase transition can be visualized unusually directly: the collective mode 
dynamics can be followed experimentally in real time and real space by imaging 
the atomic cloud. 

\end{abstract} 



\maketitle

\section{Introduction \label{sec:introduction}}

The wavefunction of a charged quantum particle moving in a magnetic field along a closed trajectory
acquires a complex phase determined by the magnetic flux penetrating the enclosed area.  An
artificial magnetic field can be designed for neutral particles by arranging for the wavefunction to
acquire the appropriate amount of phase when the particle moves in space.  This manner of creating
artificial gauge fields \cite{MagneticField,GaugeFieldReport} have been realized with photons
\cite{effectiveMagneticFieldPhotons} and in several cold atom setups, both in the continuum
\cite{Spielman_Nature2009, Spielman_NatPhys2011, Spielman_PRL2012} and in optical lattices
\cite{Bloch_exp_PRL2011, Sengstock_PRL2012, Bloch_exp_PRL2013, Ketterle_exp_PRL2013,
  Sengstock_Nature2013, Experiment, Engels_exp_PRL2015}.  In an optical lattice, a magnetic field
can be mimicked using effective complex hopping amplitudes. This is done for example using
laser-assisted tunneling.

In Ref.\ \cite{Experiment}, an effective magnetic field was created for bosons on a two-legged
ladder lattice.  In a tight-binding description, this can be described by adjusting the phases of the
hopping amplitudes so that a hopping around one plaquette increases the phase by $\phi$, which is
then regarded as the flux per plaquette.  In the Landau gauge this involves real hopping matrix
elements along the legs and complex inter-leg coupling, with the phase of successive rung phases
differing by $\phi$.  The experiment \cite{Experiment} explored the quantum phase transition between
small and large magnitudes of the rung coupling, using flux $\phi= \frac{\pi}{2}$ per plaquette.
The phase transition is between a \emph{vortex phase} at small rung hoppings, where the ground state
shows current patterns that can be described as a sequence of vortices along the ladder, and a
so-called \emph{Meissner phase} at large rung couplings.  In the Meissner phase, 
the magnetic field
is ``expelled'' from the ladder, so that the currents in the ground state run only around the
boundary of the system, i.e, chiral currents along the two legs in opposite directions.  The name
derives from analogy to the physics of type-II superconductors in a magnetic field: A type-II
superconductor expels small fields (Meissner phase) but sustains a vortex lattice above a critical
field.  In addition to the characteristic current patterns, the vortex phase is also characterized
by a modulated density profile and a degeneracy of the single-particle spectrum.

This phase transition was first studied theoretically in the context of Josephson ladders
\cite{Kardar_PRB1986,Granato_PRB1990,Orignac_Giamarchi_PRB2001}.  Several theoretical studies of
bosons in ladders subject to a magnetic field have appeared in the cold atom context, both before
and after the experimental realization of Ref.\ \cite{Experiment}.
Ref.~\cite{TheoryExperiment}, which is closely related to the experiment \cite{Experiment},  studied the
phase transition for non-interacting bosons. 
Several studies have appeared for interacting cases (Bose-Hubbard or hard-core bosons), using a
variety of methods \cite{ChaShin_PRA2011, GroundStatesFieldApproach, StrongInteraction,
  InteractionMeanField_WeiMueller_PRA2014, MottTransitionLadder, Population_imbalance,
  PhaseTransitionInteraction_Dio, YuFlach_PRE2014}.  The predominant interest in these studies,
excepting \cite{YuFlach_PRE2014}, have been phase transitions and other \emph{equilibrium}
properties in \emph{uniform} systems.


In this work, we consider the non-equilibrium dynamics of bosons in the presence of an additional
harmonic trapping potential.  Cold atomic systems are particularly suitable for tracking
non-equilibrium evolution in real time \cite{Cold-atom_reviews_expts, Naegerl_Science2009}, and this
possibility has generated an explosion of theoretical interest in the unitary dynamics of isolated
quantum many-body systems \cite{noneq_reviews}.  Now that artificial gauge fields have been realized
in several setups, non-equilibrium evolution of such systems is a promising and exciting direction
of future exploration.  Here, we present a non-equilibrium study of one such system.  We focus on
trapped systems: The experiment \cite{Experiment}, like most cold-atom experiments, was performed in
the presence of a trap.  A many-body system in a trap has collective dynamical modes which have no
analog in uniform systems, the simplest being dipole and breathing modes.  Because of the
omnipresence of trapping potentials in cold-atom experiments, dipole and breathing mode dynamics are
pervasive across the field: These modes have been intensively studied and used for diagnostic
purposes since the beginning of this field, and continue to generate interest today
\cite{Naegerl_Science2009, Esslinger_exp_PRL2003, collective_modes_early_papers,
  our_PRA_2013,KroenkeSchmelcher_BM,Bonitz_Review, Bouchoule_PRL2014, our_PRA_2015,
  STG_Astrakharchik_PRL2005, 
  STG_Chen-etal_PRA2010, br_mode_long_range_interactions, 1D_breathing_mode_recent, KohnMode}.

We study quenches (sudden changes) of the position and strength of the harmonic potential, which
generates dipole and breathing mode dynamics respectively.  For both non-interacting and interacting
systems, we identify in the spectrum the eigenstates associated with the different modes,
highlighting differences between the Meissner and vortex phases and especially the behavior at the
transition.  The eigenstates associated with the modes are (very nearly) doubly degenerate in the
vortex phase, but not in the Meissner phase.  At the phase transition, the eigenstates rearrange
their roles in order to accommodate this difference. 

Close to the phase transition we observe a slowing down of collective mode oscillations --- in the
non-interacting case the dipole frequency even vanishes at the transition.  This may be regarded as
an analog of `critical slowing down' of dynamics near classical phase transitions.  The effect found
here should be experimentally detectable through direct imaging of the oscillating size or position
of the excited cloud.  To the best of our knowledge, this is the only known visualization
possibility of a real-time slowing-down effect at a quantum phase transition.  We attribute this
effect to vanishing (or minima) of relevant energy gaps in the low energy spectrum at the critical
point.  We also show that the effect persists in interacting systems, using exact diagonalization
studies of the eigenspectrum and dynamics in multi-boson systems, and using the time-dependent
Gutzwiller approximation \cite{Gutzwiller_Jaksch}.
%
%
This phenomenon provides an unusual (real-space and real-time) manifestation of the relationship
between vanishing energy scales and diverging time scales near quantum phase transitions.  The
present results thus supplement, on the one hand, the various frequency-space consequences of this
relationship \cite{QPT_general}, and on the other hand, the indirect manifestation of
quantum-critical slowing down appearing in Kibble-Zurek-like scaling relationships for ramp (`slow
quench') dynamics \cite{noneq_reviews}.

We consider also quenches of the inter-leg hopping magnitude.  This generates breathing modes,
similar to, e.g,, interaction quenches in a trapped bosonic system without a magnetic field.  We
show that the difference of eigenstate arrangements in the two phases manifests itself in the
particular combination of eigenstates which are excited, in a manner peculiar to this system.

The paper is organized as follows.  In Sec.~\ref{sec:Hamiltonian} we introduce the Hamiltonian and
notations.
In Sec.~\ref{sec:LowSpectrum} we describe the low energy spectrum.  We identify a relevant symmetry
of the Hamiltonian and describe degeneracies of eigenstates.  For context, we contrast with a pure
one-dimensional system (e.g., single tight-binding chain) and connect excited states to collective
modes.
%
In Sec.~\ref{sec:SlowingDown} we study quenches of the trapping potential. This section contains the
most striking result of the paper, namely critical slowing down of collective modes close to the
quantum critical point.  We present and analyze exact time evolution results for systems up to four
bosons as well as mean-field (Gutzwiller) results for seven particles.  We connect the observed
dynamical slowing down to the energy spectrum.
In Sec.~\ref{sec:Kquenches} we show results for inter-leg hopping magnitude quenches. 
%
%
Sec.~\ref{sec:summary} summarizes the results and presents open problems.
In Appendix~\ref{sec:spectrum} we present the effects of the geometry of finite ladders
(boundary conditions, presence of trap) on the lower part of the energy spectrum.

\section{System and Hamiltonian \label{sec:Hamiltonian}}


The Hamiltonian we consider is
\begin{multline}
  H ~=~ - J \sum_{l} \sum_{\sigma=L,R} \left( b_{l;\sigma}^\dagger b_{l+1;\sigma} + \text{h.c.} \right)  \\
  - K \sum_l \left( e^{-i\phi l} b_{l;L}^\dagger b_{l;R} + \text{h.c.} \right) \\
  ~+~ \frac{U}{2} \sum_{l,\sigma} n_{l;\sigma} (n_{l;\sigma}-1) ~+~ \frac{\ktr}{2} \sum_{l,\sigma}
  \left( l- l_0\right)^2 n_{l;\sigma}.
\label{eq:H0}
\end{multline}
Here $b_{l;\sigma}$, $b_{l;\sigma}^\dagger$ are bosonic operators for the site on rung $l$ ($ l = 1
\ldots L$) and leg $\sigma = \text{L,R}$, $n_{l;\sigma} = b_{l;\sigma}^\dagger b_{l;\sigma}$, $\phi$
is the magnetic flux per plaquette, $J$ is the intra-leg and $K e^{{\pm}i\phi l}$ is the inter-leg
hopping matrix element, $l_0$ is the center of the trap, and $\ktr$ is the trapping strength.  We take
the trap to be centered at one of the rungs, i.e., $l_0$ is an integer.  

It is possible to use a gauge in which the complex phase is on the leg hoppings instead of the rung
hoppings, as in Refs.~\cite{Orignac_Giamarchi_PRB2001, GroundStatesFieldApproach,TheoryExperiment,
  Population_imbalance,PhaseTransitionInteraction_Dio,MottTransitionLadder}.  If the flux $\phi$ per
plaquette is the same, physical properties are unchanged.

When the density is low enough, we can approximate the cosine dispersion of a lattice particle by a
quadratic dispersion (in the leg direction).  This ``effective mass approximation'' ascribes a
continuum mass to lattice particles, $m^*=\frac{1}{2J}$, so that we can relate our trapping strength
$\ktr$ to the trapping frequency of a continuum trapping potential $\frac{1}{2}m\omega_0^2x^2$:
$\omega_0 = \sqrt{2J\ktr}$. We use this expression as the definition of the trapping frequency
$\omega_0$.

\begin{figure}[tb]
\centering
\includegraphics[width=.98\columnwidth]
{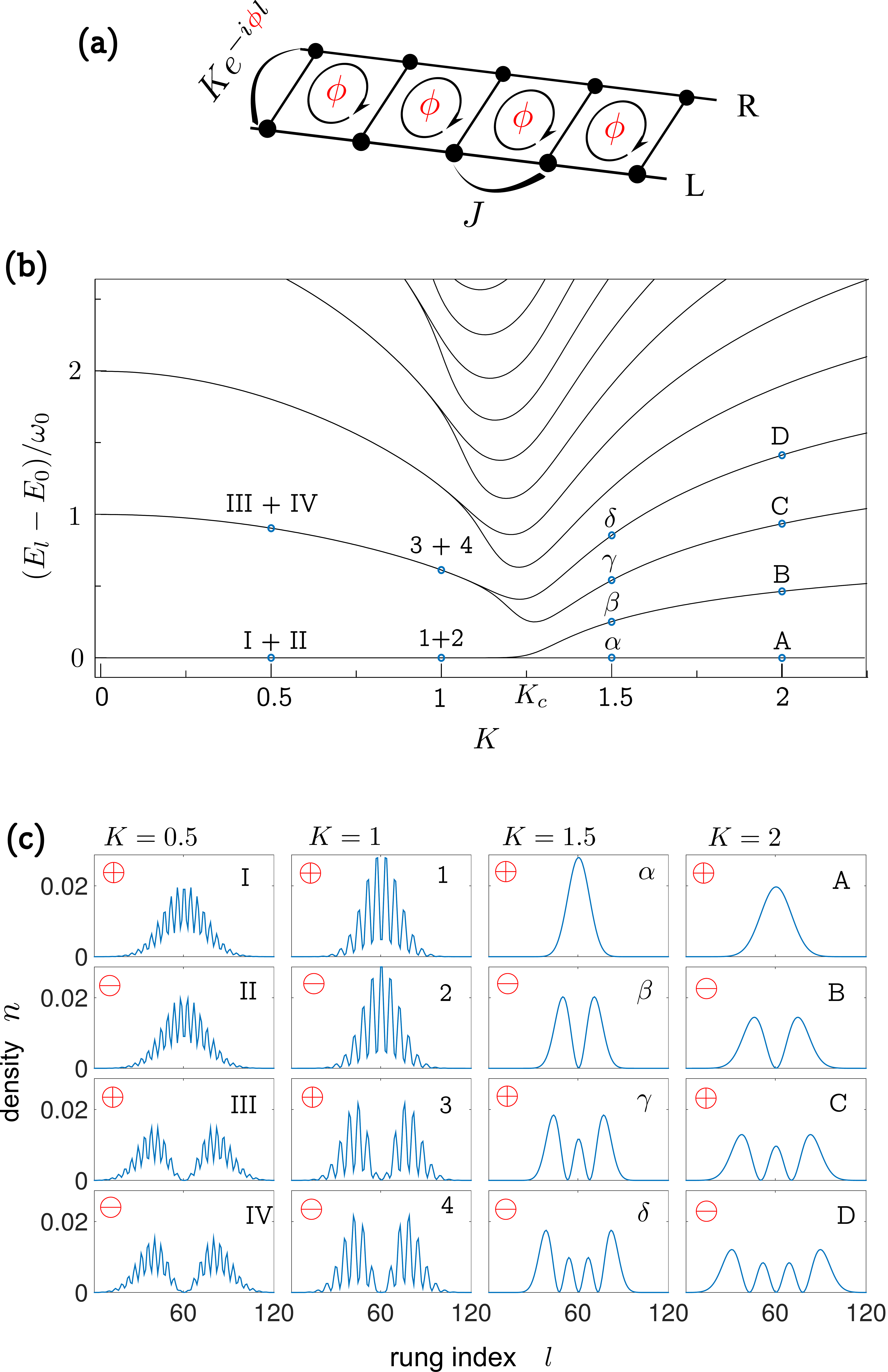}
\caption{ \label{fig:LowEnergySpectrumTrap}
%
(a) Sketch of the magnetic tight-binding ladder with complex inter-leg hoppings.  Hopping around a
plaquette increases the phase by $\phi$.
(b) Low energy excitation spectrum of the trapped magnetic ladder with 
$\phi=\frac{\pi}{2}$ for $N=1, L=500, \ktr=10^{-4}$.  The eigenstates are nearly degenerate for
$K<K_c$  (vortex phase) and non-degenerate for $K>K_c$  (Meissner phase). 
%
%
(c) Density distributions of four lowest eigenstates for $N=1, L=120, \ktr=10^{-5}$ and $K=\{0.5;1\}
< K_c$ and $K=\{1.5; 2\} > K_c$ on a leg.  For $K<K_c$ the degenerate eigenstate pairs have
identical density distributions.  The first degenerate pair have gaussian-like shape and the  second pair corresponds to 
the dipole mode.  For $K>K_c$ the states are rearranged: the second eigenstate takes the shape of a
dipole mode state and the third eigenstate becomes a breathing mode state.   The labels mark the
corresponding eigenenergy in panel (b), and the $\oplus$, $\ominus$ signs indicate parity under
pseudo-inversion. 
}
\end{figure}

Fig.~\ref{fig:LowEnergySpectrumTrap}(a) shows a sketch of the magnetic ladder system.  
%
%
The phase transition between Vortex and Meissner phases can be tuned
either by magnetic flux $\phi$ or by inter-leg hopping $K$ \cite{Experiment}. Here we fix
$\phi=\frac{\pi}{2}$ as in the experimental realization and change $K$.  The value of $K$ at the
critical point is $K_c= \sqrt{2}$ for $\phi=\frac{\pi}{2}$.  We use $\hbar=1, J=1$, measure energy
in units of $J$ and use lattice spacing $a=1$.

In the absence of a trap or interactions ($\ktr=U=0$), the phase transition can be understood by
considering the single-particle dispersion as a function of momentum
\cite{Experiment,TheoryExperiment,InteractionMeanField_WeiMueller_PRA2014}, which is well-defined
for translationally invariant systems, e.g., for infinite systems and for periodic boundary
conditions (pbc) with $L=2\pi{n}/\phi$ (integer $n$).  For $K<K_c$, the dispersion has two minima,
at nonzero momenta of opposite signs.  Thus in the vortex phase the ground state is doubly
degenerate and has finite momentum.  When translational invariance is broken, such as with open
boundary conditions (obc) or with a trap, momentum is no longer a good quantum number, but an
approximate degeneracy persists, and the density profile shows oscillations with a wavelength
corresponding to the inverse of the above-mentioned finite-momentum value.  In the Meissner phase
($K>K_c$), the dispersion has a single minimum at zero momentum; the eigenstates are not degenerate,
and the density profile in obc and trapped cases has no modulations.

\section{Lowest eigenstates of  trapped system \label{sec:LowSpectrum}}


In this section we describe the low-energy spectrum of the magnetic ladder with an additional
harmonic potential.  For harmonically trapped systems, the lowest excited states have simple
dynamical interpretations: These states correspond to various collective excitations.
Before treating the magnetic ladder, we first review the case of a pure 1D system
(either continuum or a single tight-binding chain with small fillings) in a harmonic trap.

\subsection{Review: one-dimensional system}

The 1D bosonic Hamiltonian with harmonic trapping potential is reflection-symmetric with respect to
the trap center.  The eigenstates are alternately symmetric and antisymmetric under reflection.  In
the single-particle case, these eigenstates are the familiar harmonic oscillator eigenstates, and
are characterized by an increasing number of nodes, e.g., the ground state is nodeless, and there
are one and two nodes in the first and second excited eigenfunctions respectively.

The lowest two excited states are associated respectively to dipole mode and breathing mode
excitations.  The dipole mode involves oscillations of the center-of-mass position.  It can be
excited by shifting the position of the trap minimum, i.e., a trap position quench.  The breathing
mode involves oscillations of the width of the density distribution, i.e., expansion and contraction
of the cloud.  It can be excited by a quench of either the trap strength, or (in interacting
systems) through a quench of the interaction strength.

The frequency of the collective mode oscillations is equal to the excitation energy of the
corresponding eigenstates above the ground state, i.e., $\omega_{DM} = E_{DM}- E_0 $ and
$\omega_{BM} = E_{BM}- E_0$ for the dipole and breathing mode respectively.  The dipole mode
frequency $\omega_{DM}$ is independent of interaction strength, and is locked to the trapping
frequency (Kohn's theorem \cite{KohnMode}).  In a non-interacting system, the spectrum is equally
spaced, so that the breathing mode frequency $\omega_{BM}$ is twice the trapping frequency, i.e.,
the ratio between the lowest collective modes is $\omega_{BM}/\omega_{DM} = 2$.  For an interacting
system, this ratio depends on $U$ and has a minimum at intermediate $U$, which for large systems is
$\omega_{BM}/\omega_{DM} = \sqrt{3}$ \cite{our_PRA_2013, KroenkeSchmelcher_BM,
  1D_breathing_mode_recent}.


\subsection{Pseudo-inversion symmetry of the magnetic ladder Hamiltonian} \label{sec_symmetry}

Because of the space-dependent complex phase in the rung hoppings, the magnetic ladder Hamiltonian
\eqref{eq:H0} is not symmetric under reflection about the trap center.  However, if the trap is
centered on the central rung of the ladder ($l_0=\frac{L+1}{2}$) or if the ladder is infinite, then
the Hamiltonian does possess a pseudo-inversion symmetry.  The symmetry transformation involves
reflection about the trap center and exchange of legs (reflection around rung center), together with
a space-independent phase factor.  Formally the symmetry transformation leaving the Hamiltonian
invariant is $b_{j,\mathrm{L}} \rightarrow e^{i \phi{l_0}} b_{2l_0-j,\mathrm{R}}$.
%
%
%
Each eigenstate of the Hamiltonian is either odd or even under this pseudo-inversion transformation.

Because simple reflection around the central rung is not a symmetry of the Hamiltonian, the
eigenstates are not required to have definite parity under such reflection.  There is no
straightforward correspondence between the density profile (Gaussian-like, dipole-shaped etc) and
parity, because the well-defined parity in this system corresponds to pseudo-inversion and not
reflection.
%

Note that, in order to use this symmetry, we do not need the trap center to be at the ladder center
as long as we are considering eigenstates that are localized enough that they do not feel the edges
of the chain: such eigenstates will possess well-defined parity with respect to inversion around the
center of the trap.  The symmetry is also present if the trap center is between two rungs, i.e.,
$l_0$ is half-integer.

\begin{figure*}[t]
\centering
\includegraphics[width=.9\textwidth]{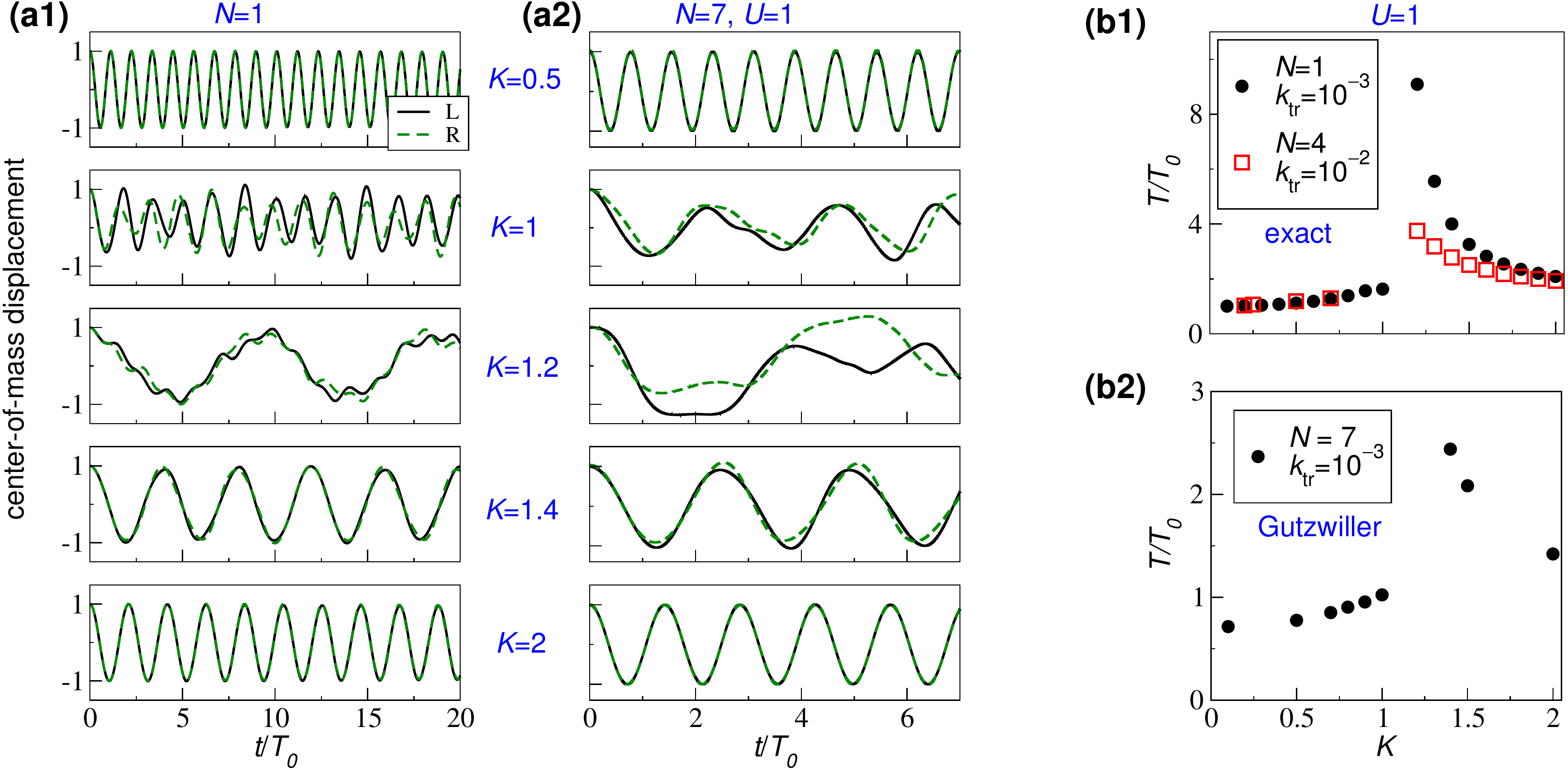}
\caption{ \label{fig:centerOfMassN1N7}
(Color online) Dipole mode oscillation after a quench (sudden shift) of trap position.
(a) Real time dynamics of center-of-mass in both legs (L,R), after  a trap shift of $d_0 = 1$ for a  
system with $L=300, \ktr=0.001$ for several $K$ values.  The slowing down effect is clearly visible:
Oscillation period is largest close to the critical point.   
(a1) Exact result for $N=1$.  
(a2) The time-evolution is calculated using Gutzwiller mean-field 
approximation for $N=7, U=1$.   
(b) Increase of dipole mode oscillation period close to the critical point at 
$K_c$. 
(b1) Exact results:  non-interacting ($N=1$) and interacting ($N=4$) systems compared.  At small $K$
and large $K$, the frequencies become identical in trap units (Kohn theorem).  (b2) Same effect is
seen with mean-field dynamics. 
}
\end{figure*}

\subsection{Single-particle eigenspectrum of trapped magnetic ladder}

We now describe the eigenspectrum of the magnetic ladder with a trap for a single particle, for
varying rung hopping $K$.  The non-interacting multi-particle spectrum can be constructed by filling
up the single-particle states.  The interacting multi-particle case will be treated briefly in a
later section.

Figure \ref{fig:LowEnergySpectrumTrap}(b) shows the lowest few states of the single-particle
spectrum, for a weak trap.  Clearly, the transition (between degenerate states at small $K$ and
non-degenerate states at large $K$) survives almost unaltered in the presence of a trap: the
low-lying states are (nearly) degenerate for small $K$ and non-degenerate for large $K$.  The point
$K_c$ where the two lowest states split is a bit smaller than in the uniform case, but $K_c
\rightarrow \sqrt{2}$ for decreasing trapping strength, $\ktr\rightarrow 0$.  The excitation energy
of excited states have a minima near $K_c$.

For weak traps ($\ktr\lesssim10^{-4}$), the eigenenergy pairs for $K<K_c$ are extremely close, the
energy differences between the lowest two states being well below double-precision accuracy (details
in Appendix~\ref{sec:spectrum}).  Thus for most purposes they can be regarded as degenerate pairs,
and we will refer to the pairs as such, even though the degeneracy is not strictly speaking exact.
Each such `degenerate' pair includes one even and one odd eigenstate under the pseudo-inversion
symmetry introduced above.

At $K \rightarrow 0 $ the two legs of the ladder are two separated tight-binding chains. Energy
eigenstates are thus truly doubly degenerate. The lowest eigenstates are $\omega_0$-spaced, in
contrast to the uniform case.
At $K \rightarrow \infty$ the magnetic ladder behaves as a single tight-binding chain with each
effective `site' composed out of a rung, resulting in an effective hopping amplitude $J^* =
\cos(\frac{\phi}{2})J = J/\sqrt{2}$ \cite{InteractionMeanField_WeiMueller_PRA2014} and effective
trapping frequency $\omega_0^* = 2^{-\frac{1}{4}}\omega_0$. In this limit the states are
non-degenerate and $\omega_0^*$-spaced.

Figure~\ref{fig:LowEnergySpectrumTrap}(c) shows plots of the density distributions of the four lowest
eigenstates at different $K$ along one leg.  The density distribution is equal on both legs.
For $K > K_c$ (lower two panels) the low energy spectrum contains non-degenerate states whose
density profiles are familiar from single-chain eigenstates: they have an increasing number of nodes
for increasing energy, and correspond to a single gaussian eigenstate, a  dipole eigenstate, a
breathing eigenstate, etc.   
For $K<K_c$, the low-lying eigenstates all have density modulations, as expected for the vortex
phase \cite{Experiment}.  The first two states (belonging to the ground state manifold) have
Gaussian-like density profiles (with additional vortex modulation). The second pair have one `node'
each in the density distribution, i.e., both have the form of a dipole eigenstate.  Thus the
gaussian, dipole, breathing,... eigenstates each come in degenerate pairs.  Each pair has one
inversion-symmetric and an inversion-antisymmetric eigenstate.

At the transition point, the states change their nature.  The ground state manifold splits into a
gaussian-shaped ground state and a dipole-mode-shaped excited state.  The other degenerate manifolds
similarly change shape: the lower of the second degenerate pair changes from a dipole profile to a
breathing mode profile having two `nodes'.

The lifting of the degeneracy of the ground state at the critical point $K_c$, and the lowering of
the other excitation energies around $K=K_c$, have intriguing dynamical consequences, detailed in
the next section. The dipole and breathing frequency decrease dramatically while approaching $K_c$.

\section{Trap quenches: Critical slowing down \label{sec:SlowingDown}}

In this section we present `critical slowing down' dynamics of the collective modes, both for
single- and multi-boson systems (for $N \geq 1$).  We excite collective modes by quenching the
trapping potential, i.e. trapping position and trapping strength for dipole and breathing modes
respectively.  We calculate time dynamics of first and second moments of the density distribution,
i.e., center-of-mass position and `width' respectively.  Close to the Meissner-vortex phase
transition in a trapped system the frequencies of collective modes are significantly suppressed,
i.e., the system slows down.

\subsection{Trap position quenches: dipole mode dynamics}

In the Figure \ref{fig:centerOfMassN1N7}(a1) we plot oscillations of the center-of-mass position
after excitation of the dipole mode for a single particle ($N=1$) at different inter-leg hopping
magnitudes $K$.  The dipole mode is excited by initializing the system in the inversion-symmetric
ground state and then shifting the center of the potential by one site, $d_0=1$.

The frequency of the dipole mode oscillation is strongly dependent on $K$.  The dipole frequency is
$\omega_0$ at $K=0$ and $\omega_0^* = 2^{-\frac{1}{4}}\omega_0$ at $K\to\infty$.  In between the
frequency has a pronounced minimum around $K_c$, as visible through the several times larger periods
at $K\approx1.2$.  The dipole frequency follows closely the excitation energy of the first excited
state for $K{\gg}K_c$ and the excitation energy of the first excited degenerate pair for $K{\ll}K_c$.
This suggests that the frequency could vanish at $K=K_c$, since the lowest two states merge as $K_c$
is approached from above.  However, in the transition region, the dynamics becomes multi-frequency,
because multiple eigenstates get excited.  (This is discussed below in the multi-particle context,
Figure \ref{fig:OverlapSpectrumN2}.)

The same slowing-down effect is observed in multi-boson systems, also in the presence of
interactions.  In the panels in Fig.~\ref{fig:centerOfMassN1N7}(a2), we plot mean-field results for
$N=7$ bosons, $U=1$, obtained using the time-dependent Gutzwiller approximation
\cite{Gutzwiller_Jaksch}.  The Gutzwiller approximation is of course not exact, but this calculation
demonstrates that the slowing down effect is generic and appears also with non-negligible
interactions.

In Figure \ref{fig:centerOfMassN1N7}(b1,b2) we show the dipole oscillation period for $N=1$, for
$N=4$ interacting bosons obtained using exact numerical diagonalization, and for $N=7$ interacting
bosons using the Gutzwiller approximation.


The dynamics shown in Fig.~\ref{fig:centerOfMassN1N7} is after a trap shift by a single lattice
site, $d_0=1$.  For stronger quenches ($d_0>1$) the time evolution is overall similar and the same
slowing down effect is seen.  For larger $d_0$, more states are excited resulting in stronger
beating effects.

Dipole excitations of harmonically trapped one-dimensional system are subject to Kohn's theorem
\cite{KohnMode}.  In the present case, only the $K=0$ and $K=\infty$ limits can be regarded as
effectively one-dimensional systems.  In accordance we see in Fig.~\ref{fig:centerOfMassN1N7}(b1)
that for small and large values of $K$, the oscillation period is equal for non-interacting and
interacting system at $U=1$, when measured in units of the trapping frequency.  There is no analogy
of Kohn's theorem for a finite $K$.

\subsection{Trap position quenches: excitation of eigenstates}

\begin{figure}
\centering  	
\includegraphics[width=.99\columnwidth]
{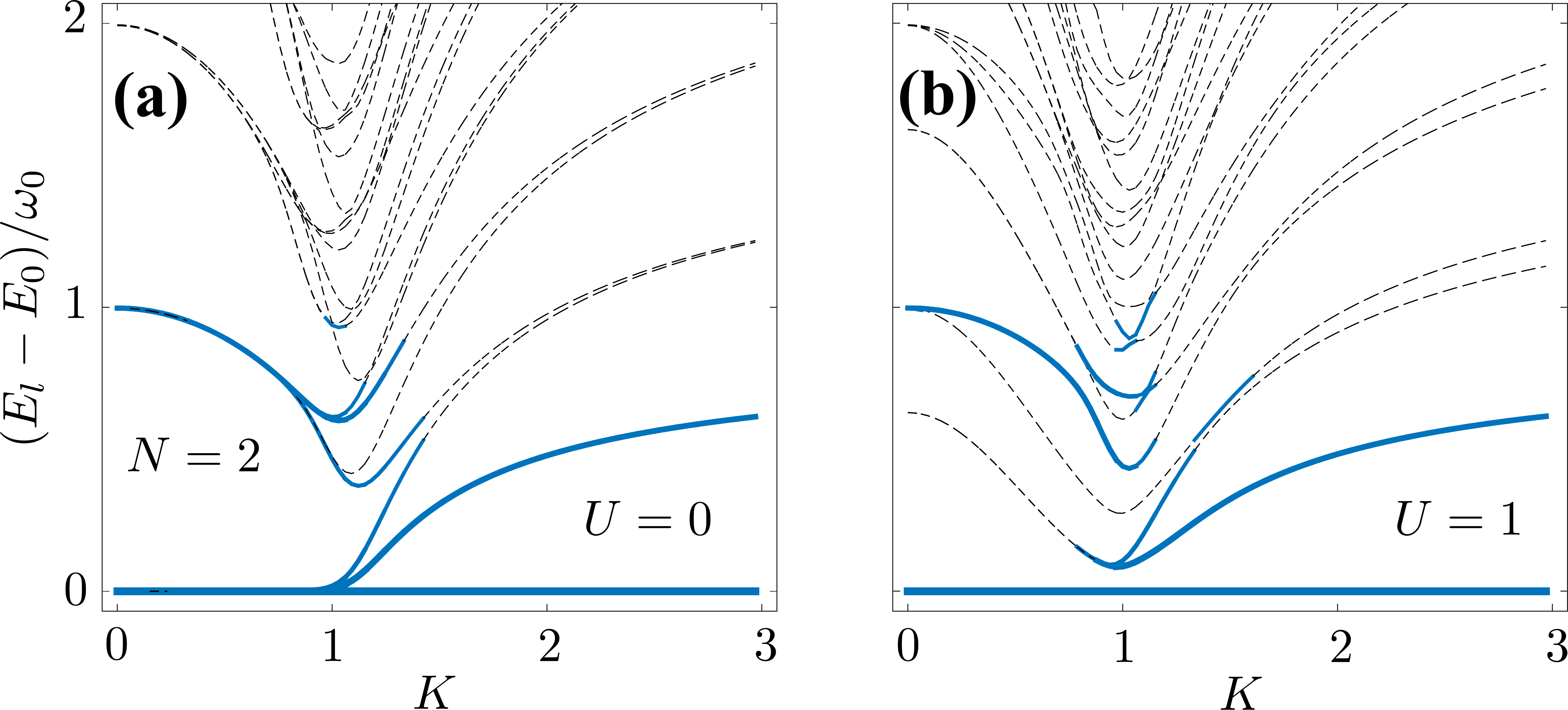}
\caption{\label{fig:OverlapSpectrumN2} (Color online) Excitation energy spectrum of the trapped
  magnetic ladder with $N=2$ bosons, $L=50$, $k=0.001$ and (a) $U=0$ and (b) $U=1$.  The line
  thicknesses logarithmically encode overlaps $|\langle \phi_0|\psi_l \rangle|^2$ of eigenstates
  with initial ground state of the Hamiltonian with a trap shift by $d_0=0.5$, i.e. after dipole
  mode excitation.  The thick, colored lines correspond to states with an overlap larger than
  $10^{-4}$. Dashed lines correspond to states with overlap smaller than $10^{-4}$.  }
\end{figure}

The dynamics after a quench is determined by the eigenstates of the final Hamiltonian which are
excited through the quench, i.e., the distribution of overlaps of the eigenstates of post-quench
Hamiltonian with the initial state, $|\langle\phi_0|\psi_l\rangle|^2$.  The energy difference
between states having substantial overlap are the dominant frequencies in the time evolution after a
quench.

Fig.~\ref{fig:OverlapSpectrumN2} shows the energy spectrum for $N=2$ bosons without and with
interactions.  The overlap distribution after a trap position quench at each value of $K$ is encoded
in the thickness of lines. 

Without interactions ($U=0$, Fig.~\ref{fig:OverlapSpectrumN2}(a)), the $N>1$ spectrum has the same
qualitative features as for $N=1$, except for additional degeneracies at $K<K_c$ and at large
$K{\gg}K_c$, and visibly larger number of states around $K\approx{K_c}$. The larger number of states
and greater degeneracies can be understood as the number of possibilities for multiple bosons to
occupy the single-particle states.  Away from the critical region, the effect of a trap position
quench is to excite mainly the new ground state and the first excited state (dipole eigenstate), so
that the lowest energy gap sets the dipole frequency.  In the $K\approx K_c$ region, several states
are seen to be excited, leading to multi-frequency dynamics reported already in
Fig.~\ref{fig:centerOfMassN1N7}(a).  

Interactions lift degeneracies in the energy spectrum, Fig.~\ref{fig:OverlapSpectrumN2}(b).  For
$K<K_c$, this leads to the situation, perhaps surprising at first sight, that the dipole-mode
eigenstate is not the lowest excited state. This does not occur in pure one-dimensional
(single-chain) systems \cite{our_PRA_2013, our_PRA_2015} and is a novel feature of ladder
geometries.
The excitation energies of the lowest excited states have a minimum at $K\approx K_c$ but there is
no proper gap closing as one approaches from the $K>K_c$ side, in contrast to the $U=0$ case.  As in
the $U=0$ case, the overlap distribution after a trap quench is dominated by the ground state and
dipole eigenstate away from the critical region, while there are multiple eigenstates excited in the
$K\approx K_c$ region.  For $K<K_c$, the excited states which appear below the dipole-mode state are
not excited.  The dipole-mode eigenenergy difference with the ground state energy is $\omega_0$ at
$K=0$ and $\omega_0^* = 2^{-\frac{1}{4}}\omega_0$ at $K\to\infty$, just as in the $U=0$ case, in
accordance with Kohn's theorem according to which this energy difference should be $U$-independent.

\begin{figure}[t]
\centering
\includegraphics[width=.98\columnwidth]{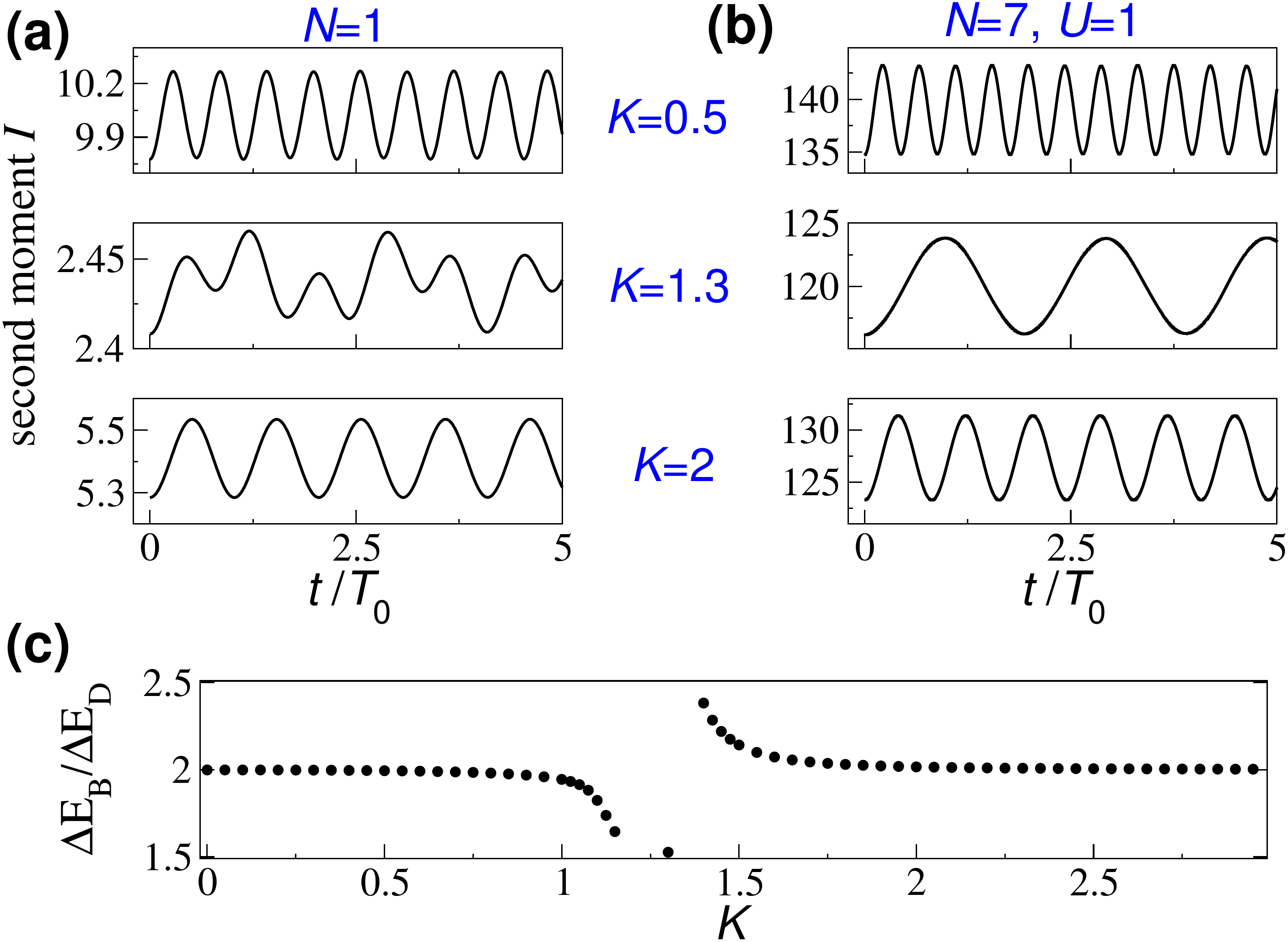}
\caption{ \label{fig:SecondMomentN1N7} 
(Color online) 
Breathing mode oscillation after a trap quench for various $K$, shown using the second moment of the
density distribution for $L= 300$, $\ktr=0.00105 \rightarrow 0.001$. 
(a) Exact result for $N=1$.
(b) Gutzwiller mean-field approximation for $N=7, U=1$. 
(c) Ratio of the breathing and dipole mode frequencies for $N=1$, $L=300$, $\ktr = 0.0001$.  The two
excitations are obtained using a  $d_0=1$ quench (dipole mode) and a $\ktr=0.00011 \to 0.0001$
quench (breathing mode).  
}
\end{figure}

\subsection{Trap strength quenches: breathing mode}

A similar slowing-down effect is visible for the breathing mode oscillations, in
Fig.~\ref{fig:SecondMomentN1N7}(a,b).   The frequency reduction is smaller compared to the dipole
mode case. 

In Fig.~\ref{fig:SecondMomentN1N7}(c) we plot the ratio of the energy gaps corresponding to
breathing and dipole mode oscillations, calculated from the energy spectrum as a function of
inter-leg hopping $K$.  The energy gap used is that between the ground state and the next most
strongly excited state in a trap strength quench (for the breathing mode) and in a trap position
quench (for the dipole mode).  Away from the critical point, the ratio is equal to two, as expected
for a single particle.  Close to the critical point, the behavior is completely different from
non-magnetic system; the divergence-like feature signals that the dipole frequency nearly vanishes
in this region.

\section{Inter-leg hopping quenches \label{sec:Kquenches}}

\begin{figure}[t]
\centering
\includegraphics[width=.98\columnwidth]
{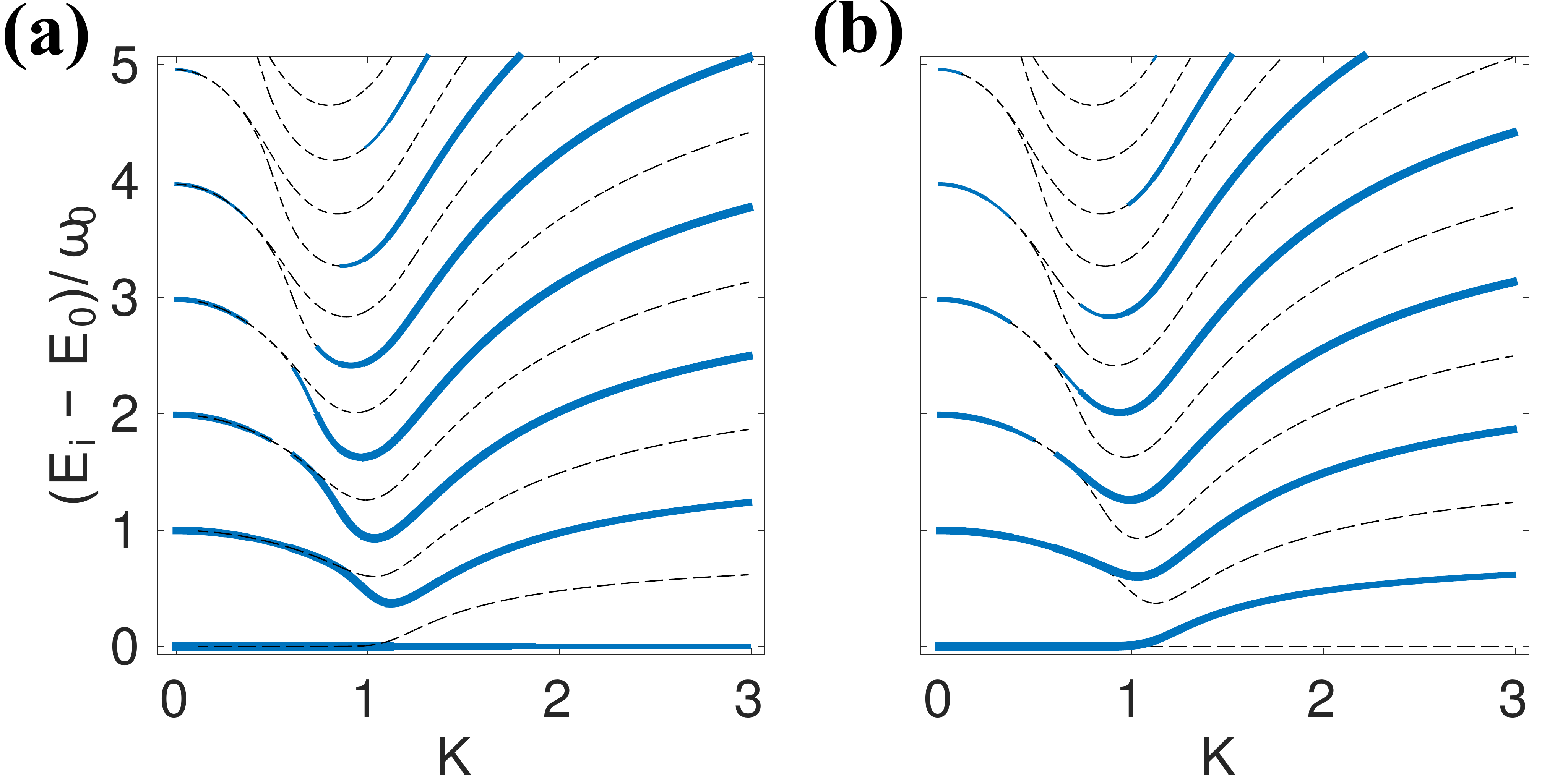}
\caption{ \label{fig:OverlapKquench} (Color online) Excitations after a 
$K$-quench.  We show  overlaps of the eigenstates with an initial state, encoded 
in the thickness of the solid lines. The initial state is one state of the 
ground state manifold with $K_i=0.5$, $N=1$, $L=200$, $\ktr=0.001$ with (a) 
and (b) odd parity under inversion. Only low energy states with the same
  symmetry as the initial state are excited.  }
\end{figure}

In this section we treat quenches of the inter-leg hopping magnitude $K$.  In
Fig.~\ref{fig:OverlapKquench}, we fix the initial $K_i$ to be in the vortex phase, $K_i<K_c$, and
show overlaps with eigenstates of the final Hamiltonian for various $K_f$.  We choose the initial
state to be one of the two states of the ground state manifold for $K_i < K_c$, i.e. either
symmetric or anti-symmetric with respect to the pseudo-inversion transformation introduced in
subsection \ref{sec_symmetry}.  Note that both these states have gaussian-like density profiles with
additional density modulations (Fig.~\ref{fig:LowEnergySpectrumTrap}(c), states I,II,1,2).

For $K_f < K_c$, the quench chooses from each degenerate pair the state with the same symmetry as
the initial state, and one of each degenerate pair has nonzero overlap.  The highest overlap is with
the ground state, and the overlaps get progressively smaller with increasing eigenenergy.  For $K_f
> K_c$, alternate eigenstates have the same symmetry, so that the nonzero overlaps are either only
with odd states, or only with even states.  The two cases are shown in the two panels of Figure
\ref{fig:OverlapKquench}.  The situation on the right panel may appear counter-intuitive, as the
eigenstates excited have density profiles that appear odd, which for a pure one-dimensional system
one would not expect starting from a state with a gaussian profile.  However, the more complicated
symmetry of the present case allows odd-symmetry states to have gaussian density profile and
even-symmetry states to have dipole-like density profile.  

The temporal dynamics after a $K$ quench involves complicated multi-frequency oscillations including
both breathing and dipole-like components in the densities of both legs (not shown).  The total
center-of-mass stays fixed but there are out-of-phase dipole oscillations in the two legs.

\section{Summary and Discussion \label{sec:summary}}

In this work, we have provided a theoretical study of some non-equilibrium dynamics of bosons in the
magnetic ladder system recently realized in experiment \cite{Experiment}, focusing in particular on
effects due to the presence of a harmonic trapping potential, which is almost universally present in
such experiments.  Previous theory studies have focused heavily on the uniform system and
equilibrium properties.  Other than this work, non-equilibrium non-uniform dynamics was also studied
in Ref.\ \cite{YuFlach_PRE2014}, where a nonlinear version of the system was studied in the presence
of disorder.  This may be regarded as a mean-field approximation to a many-boson system with on-site
interactions.

By studying trap quenches, we have presented the dynamics and spectral structures related to dipole
modes and breathing modes.  The most striking result is a slowing-down effect in these modes near
the critical point, an effect that should be readily observable in the experimental setup.  We are
unaware of any other example of a possibility to visualize quantum critical dynamics in real-space
oscillation phenomena.
Critical slowing down in real-time dynamics is of course well-known for classical phase transitions
\cite{ChaikinLubensky_book}, where the phenomenon can be dramatically visualized, e.g., through the
response of a liquid sample near a liquid-gas transition point after being shaken.


We have stressed in particular the connection of specific dynamical phenomena to structures in the
eigenspectrum.  This correspondence, and the fact that noteworthy spectral structures can rise to
unusual dynamical phenomena, is increasingly appreciated and explicitly studied in the recent
non-equilibrium literature, e.g., in the context of self-trapping and related phenomena in
Bose-Hubbard dimers and ladders \cite{BoseHubbardDimer_various,BHladder_Bloch_expt_theory}, in the
study of (repulsive) binding of lattice particles \cite{bound_clusters_itinerant}, collective modes
of trapped systems \cite{our_PRA_2013, our_PRA_2015}, dynamics in spin chains
\cite{XXZ_spectrum_dynamics}, and various other phenomena \cite{MazzaFabrizio_PRB12,
  ChancellorPetriHaas_PRB13, Wimberger_PRA13}.  Our results connecting critical slowing down of
collective modes to minima in the excitation gaps (or splitting of near-degenerate eigenstate pairs)
can be seen as an addition to this theme.
In addition, in this work we have examined the eigenstates of the final Hamiltonian that are excited
in a quench, i.e., the distribution of overlaps of the final eigenstates with the initial state as a
function of energy.  This overlap distribution excited in a quench, and its continuous version, the
work distribution, has been the focus of much recent attention
\cite{LSantos_workdistr, work_distribution,
  work_dist_condmat_models}.

A peculiar feature of the present system is that gaussian-like and dipole-like shapes do not
necessary correspond to even and odd parity respectively, in contrast to the experience obtained
from studying purely one-dimensional trapped systems or the harmonic oscillator eigenstates.  The
reason is that the conserved parity in this case does not relate to a simple reflection around the
center rung, but rather a pseudo-inversion involving reflections along both leg and rung directions
(plus a possible phase shift).  This allows somewhat unexpected situations like the first two
(near-degenerate) eigenstates both having gaussian-like profile, the second pair having dipole-like
profile, etc., in the vortex phase.  At the transition, the states rearrange and change nature, and
in the Meissner phase we get a more familiar sequence of a single ground state with gaussian-like
density profile, only the first excited eigenstate with dipole-like profile, and so on.

We have also shown how interactions modify the spectrum for a multi-boson trapped system.
Interactions lead to the peculiar effect that, in the Meissner phase, the dipole eigenstates are no
longer the first excited state or first excited pair; instead there are eigenstates between the
ground state and the dipole eigenstate.  This is a situation that does not occur in pure 1D chains
\cite{our_PRA_2013, our_PRA_2015, KroenkeSchmelcher_BM}.  

The present work raises a number of open questions.  The very nearly perfect degeneracy in trapped
finite systems remains unexplained.  As we show in the Appendix (Fig.~\ref{fig:OverlapSpectrumN2}),
open boundary conditions cause degeneracy-breaking gaps in the Meissner phase that are many orders
of magnitude larger than in even more tightly confined harmonic-trapped cases.  Analytic solutions
for the obc and trapped cases would be helpful in this regard but are currently not known.
Physically, one can speculate that harmonic trapping is a less violent manner of breaking
translation symmetry than open ends of the ladder.  More generally, this is one of the first studies
of non-equilibrium dynamics in the magnetic ladder, treating the most obvious quench protocols and
the corresponding parts of the eigenspectrum.  We expect many other interesting dynamical phenomena,
e.g,. corresponding to aspects of the eigenspectrum that have not yet been explored, or involving
interesting types of current circulation dynamics.

\begin{acknowledgments}

  We thank I.~Bloch and M.~Vojta for discussions.

\end{acknowledgments}

\appendix

\section{Geometry \label{sec:spectrum}}

\begin{figure}[t]
\centering
\includegraphics[width=.98\columnwidth]{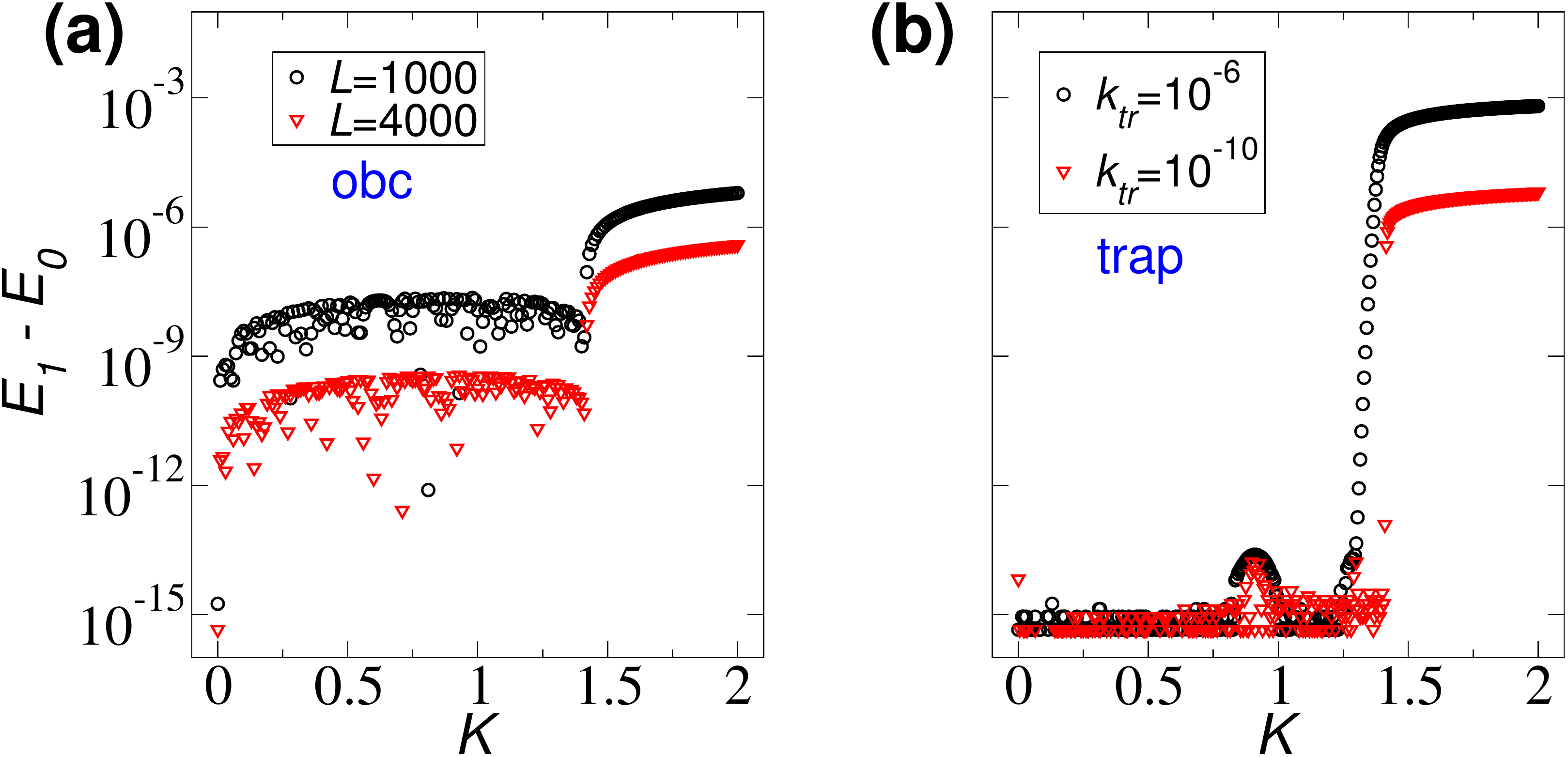}
\caption{ \label{fig:energyGapTrap_obc}
(Color online) Energy gap between lowest two single-particle eigenstates, obtained using
double-precision numerics.  
(a) Open boundary conditions, various sizes. The gap in the vortex phase is seen to be numerically
significant, ${\gg}10^{-15}$.  
(b) Trapped system with various $\ktr$.  Data is obtained with $L=4000$, for which the trapped cloud
does not reach the edges up to numerical accuracy.  The gap is seen to reach ``numerical zero'',
$\sim10^{-15}$.  }
\end{figure}

In this work we concentrate on trapped systems, in contrast to most of the literature on the
magnetic ladder systems which focus on uniform systems.  To provide context, we describe here some
effects of the geometry of the system, by contrasting spectral features of periodic, open-boundary
and trapped systems.

\paragraph*{Periodic boundary conditions ---}

When $L$ is a multiple of $2\pi/\phi=4$, a system with pbc is uniform, and momentum along the leg
direction is a good quantum number.  In such a case diagonalization can be done analytically in the
absence of interactions, by Bogoliubov rotation in momentum space.  The resulting
single-particle spectrum has two branches, which we find to be
\begin{equation}
E_{(1)} = -2JA_q-2KC_q \, , \quad E_{(2)} = -2JB_q+2KC_q
\end{equation}
with 
\[
A_q = \cos{q}\sin^2\Theta + \cos(q+\phi)\cos^2\Theta  ,
\] 
\[
B_q = \cos(q+\phi)\sin^2\Theta + \cos{q}\cos^2\Theta  ,
\]  
\[
C_q = \sin\Theta \cos\Theta  ,
\] 
and 
\[
\Theta = \frac{1}{2}\arctan
	  \left( \frac{K}{J( \cos(q+\phi) - \cos(q) )} \right).
\]
Momenta take the values $q = \frac{2\pi}{L}n$, with an integer $n=-L/2, -L/2+1, 
\dots, L/2-1$.

When $L$ is an even multiple of 4, i.e., a multiple of 8, momenta values can be 
shifted by $\phi/2$: $k = q - \phi/2$ recovering 
the expression appearing previously in the literature \cite{TheoryExperiment,
  InteractionMeanField_WeiMueller_PRA2014, MottTransitionLadder}:
\begin{equation}
E_{\pm} = -2J\cos k \cos\frac{\phi}{2} \pm \sqrt{4J^2 \sin^2k 
\sin^2\frac{\phi}{2} + K^2}.
\label{eq:energyL8n}
\end{equation}

The spectrum contains two minima for $K<K_c$ (hence the exact degeneracy) and 
one minimum for $K>K_c$ (hence the single ground state).
Note that this is correct \emph{only} for $L=8n$ with integer $n$.  
For $L=4n$ with an odd $n$ energy spectrum contains doubly degenerate states 
fro all $K$.

When $L$ is not a multiple of 4, we are unaware of an analytical solution for 
pbc without any twist in the boundary condition.  Of course, expression 
\eqref{eq:energyL8n} was previously used with large sizes in mind, where the 
boundary conditions become insignificant.  Our purpose here is to present 
explicitly the effects of geometry.

When momentum is well-defined ($L=4n$), almost all states are degenerate.  The only exceptions are
the $k=0$ states for even $n$ ($L=8n$).  At small $K$, the full spectrum (not shown) shows many true
level crossings between such degenerate pairs.

\paragraph*{Open boundary conditions ---}

With obc, momentum is not a good quantum number.  Level crossings are therefore avoided, so that the
full spectrum looks visually quite different from the pbc case.  Of course, for large enough system
sizes, the low-energy part of the spectrum looks similar to the pbc case.  For $K<K_c$, the energy
levels are grouped in pairs, each such pair undergo a succession of mutual avoided level crossings
as a function of $K$.  With increasing $L$, the gap within each pair decreases, as shown in
Fig.~\ref{fig:energyGapTrap_obc}(a).  This is consistent with the obc case in the large-$L$ limit
becoming similar to the pbc case (where the $K<K_c$ pairs are degenerate).

\paragraph*{System with harmonic trap ---}

The energy spectrum of the trapped system is similar to a system with obc; momentum is not a good
quantum number.  However, the trapped system displays extremely small gaps between the degenerate
pairs for $K<K_c$.  In Fig.~\ref{fig:energyGapTrap_obc}(b) we show the lowest gap as obtained from
double-precision computations.  The pairs are `degenerate' up to standard numerical accuracy already
for $\ktr\approx10^{-4}$.  Numerical diagonalization with higher precision shows that the degeneracy
is not exact and there is a minute but finite gap for finite $\ktr$.  At present a detailed
explanation of this ultra-small gap is lacking, since we do not have analytical solutions for the
trapped or obc cases.


\end{document}